\documentclass{ifacconf}

\usepackage{filecontents}

\makeatletter
\let\old@ssect\@ssect 
\makeatother

\usepackage{algpseudocode}
\usepackage{algorithm}
\usepackage{amsmath,amssymb}
\usepackage{graphicx}      
\usepackage{epstopdf}
\usepackage{natbib}        

\newtheorem{remark}{Remark}[section]

\newtheorem{example}{Example}[section]

\usepackage{color, hyperref}
\definecolor{darkblue}{rgb}{0.0,0.0,0.6}
\hypersetup{colorlinks,breaklinks,linkcolor=darkblue,urlcolor=darkblue,anchorcolor=darkblue,citecolor=darkblue}
\makeatletter
\def\@ssect#1#2#3#4#5#6{%
	\NR@gettitle{#6}
	\old@ssect{#1}{#2}{#3}{#4}{#5}{#6}
}
\makeatother
\begin{document}
	\begin{frontmatter}
		
		\title{On Data-Driven Log-Optimal Portfolio: \\A Sliding Window Approach\thanksref{footnoteinfo}} 
		
		\thanks[footnoteinfo]{This paper is partially supported by the Ministry of Science and Technology~(MOST), Taiwan, R.O.C. under Grant:  MOST--110--2222--E--007--005.}
		
		\author[First]{Pei-Ting Wang} and \author[First]{Chung-Han Hsieh}
		
		\address[First]{Department of Quantitative Finance, National Tsing Hua University, Hsinchu 30044, Taiwan R.O.C. (e-mail: \href{mailto: mochikimosu0426@gmail.com}{mochikimosu0426@gmail.com}, \href{ch.hsieh@mx.nthu.edu.tw}{ch.hsieh@mx.nthu.edu.tw}). }

		\begin{abstract}    
			In this paper, we propose a data-driven  \textit{sliding window} approach to solve a log-optimal portfolio problem.
			In contrast to many of the existing papers, this approach leads to a trading strategy with time-varying portfolio weights rather than fixed constant weights. 
			We show, by conducting various empirical studies, that the approach possesses a superior trading performance to the classical log-optimal portfolio in the sense of having a higher cumulative rate of returns.
		\end{abstract}
		
		\begin{keyword}
			Stochastic Systems, Data-Driven Approach, Financial Engineering, Empirical Portfolio Optimization
		\end{keyword}
		
	\end{frontmatter}
	
	\section{Introduction}
	\label{SECTION: INTRODUCTION}
	The take-off point for this paper is the so-called \textit{log-optimal portfolio}, which is any portfolio that maximizes the Expected Logarithmic Growth~(ELG) of a trader's wealth.
	This ELG maximization idea was originated from~\cite{Kelly_1956} to solve repeated coin-flipping gambling problems.  
	Since then, various ramifications and extensions along this line are studied extensively; e.g., see  \cite{algoet1988asymptotic, rotando1992kelly, cover2006elements, thorp2006kelly}.
	Specifically, the objective for solving the classical ELG problem is to seek a portfolio with weight~$K$ of its account value that maximizes the ELG at the terminal stage.
	Some good and bad properties are studied in \cite{maclean2010good}.
	A rather comprehensive survey of the Kelly-based approach can be found in \cite{maclean2011kelly} and the references therein.
	While many of the existing papers contributed to Kelly's problem and its application to stock trading; e.g., see~{\cite{rotando1992kelly, algoet1988asymptotic, cover2006elements, lo2018growth, kuhn2010analysis, thorp2006kelly, hsieh2016kelly}}, the resulting optimal weights $K$ are typically time-invariant.

	According to \cite{cover2006elements, cornuejols2006optimization}, it is known that if the returns are independent and identically distributed~(IID) and known perfectly to the trader, then the constant weight~$K$ is~optimal.
	However, in practice, the returns are typically neither~IID nor perfectly known to the trader; see \cite{fama2021market} and \cite{luenberger2013investment}. 
	To this end, various approaches are proposed to solve the log-optimal portfolio problem by relaxing the IID assumptions; e.g., see \cite{cover1991universal} for proposing a \textit{universal portfolio} with unknown return distributions and \cite{rujeerapaiboon2018risk, rujeerapaiboon2016robust, sun2018distributional} for solving various robust Kelly optimal portfolio problems.
	In contrast, this paper,  under somewhat weak assumptions on returns, aimed at studying the data-driven log-optimal portfolio problem via a \textit{sliding window approach}.
	Unlike many of the papers mentioned above that considering a portfolio with a \textit{fixed constant} weight; see \cite{algoet1988asymptotic,  thorp2006kelly, nekrasov2014kelly, hsieh2018rebalancing, hsieh2021necessary, hsieh2022generalization}, our approach leads to a time-varying portfolio weight, which generalizes the case with a fixed constant weight.
	To close this brief introduction, we also mention some related work here; e.g., see \cite{park2007we} for a good survey of the profitability of technical analysis, \cite{wu2022optiontrading} for an interesting application of using Kelly-based approach in options trading, \cite{hsieh2020feedback} for analyzing log-optimal fraction using Taylor-based approximation approach.

	\subsection{Plans for Section to Follow}
	In Section~\ref{section: Problem Formulation}, we provide some preliminaries and formulate a data-driven log-optimal portfolio problem in a discrete-time setting.
	Then, in Section~\ref{Section: Sliding Window Approach}, a sliding window algorithm is provided to solve the portfolio problem.
	Subsequently, in Section~\ref{section: Empirical Sutdies}, with the sliding window approach, we provide various empirical studies using historical stock price data.
	Lastly, in Section~\ref{section: Concluding Remarks}, some concluding remarks and future research directions are discussed.
	
	\section{Problem Formulation}
	\label{section: Problem Formulation}
	Fix an integer $N>1$. 
	For stages~$k=0,1,\dots, N-1$, we consider a trader who is forming a portfolio consisting of~$m \geq 2$ assets.
	If an asset is riskless, its return is deterministic and is treated as a degenerate random variable with value $X(k)=r\geq 0$ for all $k$ with probability one.
	Alternatively, if Asset~$i$ is a stock whose price at time~$k$ is~$S_i(k)>0$, then its return is
	\[
	X_i(k) = \frac{S_i(k+1) - S_i(k)}{S_i(k)}.
	\]
	In the sequel,  we assume that the return vectors
	$$
	X(k):=\left[X_1(k) \; X_2(k)\;  \cdots \;X_m(k)\right]^T.
	$$
	The return vector $X(k)$ is drawn according to an   \textit{unknown} but identically distributed distribution which is supported on $J$ distinct points.
	For~$k=0,1,\dots,N-1$, the corresponding joint probability mass function can be estimated~by
	\[
	P(X_1(k) = x_1^j, \dots, X_m(k) = x_m^j) = P(X(k)=x^j)=p_j
	\]
	for $j=1,2,\dots, J$
	with $p_j \geq 0$ and $\sum_{j=1}^J p_j = 1.$
	We also assume that the returns satisfy
	$
	X_{\min,i}  \leq X_i(k) \leq X_{\max,i}
	$
	with known bounds above and with $X_{\max,i}$ being finite and~\mbox{$X_{\min,i} > -1$}.

	\subsection{Portfolio Weight and Account Value Dynamics}
	For $k=0,1,\dots$, let $V(k)$ be the trader's account value at stage $k$.
	Now,  for~$i=1, 2, \dots, m$, take
	$
	0  \leq K_i \leq 1,
	$
	which represents a weight of the portfolio allocated to the~$i$th asset.
	In the sequel, we require that the trade is \textit{long-only} and \textit{cash-financed}. 
	Specifically, using a shorthand notation~$
	K:=[K_1\;K_2\; \cdots \; K_m]^T,
	$
	we require that the weight must satisfy the classical \textit{unit simplex} constraint;~i.e.,
	\[
	K \in \mathcal{K} := 
	\left\{ K \in \mathbb{R}^{m}: K_i \geq 0  \text{ for all } \ i,\ \sum_{i=1}^m K_i = 1 \right\}.
	\]
	
	Now, at stage $k=0,$ we begin with the initial account value~$V(0):=V_0 > 0 $. 
	The associated account value dynamics can be described as the following stochastic recursive equation
	\begin{align*}
		V(k+1) 
		&= (1+K^TX(k)) V(k).
	\end{align*}
	The \textit{per-period log-return} is defined as
	\[
	g(k):=\log \frac{V(k+1)}{V(k)}
	\]
	which is used as the objective of the maximization in the sections to follow. 
	
	\subsection{Data-Driven Log-Optimal Portfolio Problem}
	To incorporate with available data, we modify the problem and consider the data-driven  log-optimal portfolio problem:
	\begin{align}
		\max_{K \in \mathcal{K}} \ \mathbb{E}\left[ g(k) \right] 
		& = \max_{K \in \mathcal{K}} \ \mathbb{E}[\ \log (1+K^T X(k))\ ] \nonumber\\
		& = \max_{K \in \mathcal{K}} \ \sum_{j=1}^J p_j  \log (1+K^T x^j). \label{problem: log-optimal sliding}
	\end{align}
	The problem above is readily verified as a concave program; i.e., a maximization problem  with a concave objective and a convex constraint set $\cal K$; see \cite{boyd2004convex}.
	In this paper, we will use a modeling language called
	CVXPY (using Python language); see~\cite{diamond2016cvxpy}. CVXPY enable us to solve the problem in a very efficient manner.
	
	\begin{remark}
		To close this section, it is worth mentioning that, unlike many of the existing papers in solving log-optimal portfolio using fixed constant weight, our approach seeks to find a portfolio weight~$K \in \mathcal{K}$  by solving the latest optimal portfolio problem with sliding window sizes~$M \geq 1$.
		The resulting optimal weight~$K \in \mathcal{K}$ will be applied to the successive stage $k+1$. Then we ``slide" the window and resolve the problem. The procedure is repeated until the terminal stage.
	\end{remark}


	\section{Sliding Window Algorithm} \label{Section: Sliding Window Approach}
	In this section,   we are ready to introduce our sliding window approach aimed at solving the data-driven log-optimal portfolio problem stated in Section~\ref{section: Problem Formulation}.
	Specifically, for each~$k=0,1,\dots$, we rebalance our portfolio using a sliding window with sizes $M$ days to obtain the latest optimal weight.
	
	Using $s_i(k)$ to denote the realized prices for Asset~$i$ at stage $k$, the associated realized returns $x_i(k)$ satisfy
	\begin{align} \label{eq: x_i realized returns}
		x_i(k) := \frac{s_i(k+1) - s_i(k)}{s_i(k)}
	\end{align}
	for $i=1,2,\dots, m$.
	Let
	$$
	x(k):=\left[x_1(k) \; x_2(k)\;  \cdots \;x_m(k)\right]^T.
	$$
	Now sitting at $k=k_0$, the sliding window approach is aimed at obtaining an optimal $K^*$ that will be applied to the next stage $k_0+1$.
	To this end, we proceed as follows.
	We begin by selecting a collection of realized returns~$\{x(j)\}_{j=k_0-M}^{k_0-1}$, which is obtained in~$M$ past stages. 
	Subsequently, we solve the  maximization problem to obtain the $K^*$. The detailed algorithm is summarized below.

	\newcommand{\INDSTATE}[1][1]{\STATE\hspace{#1\algorithmicindent}}
	\begin{algorithm}[h!] 
		\caption{Data-Driven Sliding Window Algorithm}\label{algorithm: sliding window}
		\begin{algorithmic}[1]
			\Require
			Consider $m \geq 2$ assets, 
			stock prices $\{s_i(j)\}_{k=0}^{N-1}$ for $i=1,2,\dots, m$, 
			and sliding window size $M$.
			\Ensure 
			Optimal portfolio weight $K^*$ for each stage
			\State Compute realized returns $\{x_i(k)\}_{k=0}^{N-1}$ with $x_i(k)$ defined in Equation~(\ref{eq: x_i realized returns}). 
			\If {$k \geq M$}
			\State Choose $\{x(j)\}_{j=k-M}^{k-1}$ from $\{x(k)\}_{k=0}^{N-1}$ with joint probability $p_j:=1/M$ for all $j=k-M,\dots, k-1$.
			\State Solve the maximization problem 
			$$
			\max_{K \in \mathcal{K}} \ \frac{1}{M} \sum\limits_{j=k-M}^{k-1} \ \log (1+K^T  x(j)  ) .
			$$
			\State Having obtained optimal $K^*(k):=K^*$, we apply it at stage~$k+1$. Set $k=k+1$ then back to Step 2.
			\EndIf
		\end{algorithmic}
	\end{algorithm}
	\section{Empirical Studies} \label{section: Empirical Sutdies}
	To illustrate our sliding window approach, two empirical studies are considered: The first one is a three-asset portfolio using daily prices and the second one is the same portfolio using the flipped upside down prices. 	Consistent with the existing literature; e.g.,~\cite{bodie2018investments}, to compare the trading performance, some typical metrics are used as follows. 
	
	\subsection{Performance Metrics}
	For $k=0,1,\dots, N-1$, the first performance metric to be used in the analysis to follow is the \textit{portfolio realized return} in period~$k$. That is,  
	\[
	R^p(k) := \frac{V(k+1) - V(k)}{V(k)} .
	\]
	The (realized) cumulative return up to stage $k=N$ is given by $(V(N) - V(0) )/ V(0)$ and the log-growth rate is the logarithm of the realized cumulative return; i.e.,~$\log (V(N)/V(0))$.
	The \textit{excess return} denoted 
	$$R^p(k) := R^p(k) - r_f$$ where~$r_f$ is the \textit{risk-free rate}.
	The \textit{realized (per-period) Sharpe ratio}, denoted by $SR$, of the portfolio is the average of the excess returns $\overline{R^p}$ over the standard deviation of the excess returns~$\sigma$; i.e., 
	$$
	SR:= \frac{\overline{R^p} - r_f}{ \sigma}
	$$
	and the \textit{N-period realized Sharpe ratio} can be approximated by $\sqrt{N}\cdot SR$; see \cite{lo2002statistics}.
	Lastly, other than standard deviation, to scrutinize the downside risks over multi-period trading performance, we include the \textit{maximum percentage drawdown} as an alternative risk metric;~i.e.,
	\[
	d^*:= \max_{0\leq \ell < k \leq N} \frac{V(\ell) - V(k)}{V(\ell)}.
	\]
	Lastly, we also report the running times of the sliding window approach. 
	For example, with $m=3$ assets, we generate  $10,000$ of sample paths using the historical daily price data with window sizes $M =10$. 
	On a laptop with~$2.5$~GHz with $4$GB RAM, the data-driven log-optimal problem can be solved  about $3.18$ seconds, which is in the same scale of solving classical log-optimal portfolio.\footnote{The required running times justifies the sliding window approach with daily price data since the time required to solve the problem is less than a day.}

	\medskip
	\begin{example}[Three Assets Portfolio] \label{example: three asset portfolio}
		Consider a portfolio consisting of three underlying assets: The first asset is Vanguard Total World Stock~(Ticker:~VT), the second asset is Vanguard Total Bond Market ETF (Ticker: BND), and the third asset is Vanguard Total International Bond~ETF~(Ticker:~BNDX).\footnote{It is worth mentioning that these three assets (VT, BND, and BNDX) forms a well-diversified portfolio.} 
		Assuming $r_f:=0$ and a two-year duration from February~14,~2018 to February~14,~2020, we solve the log-optimal portfolio problem to obtain the optimal weight $K^*$. 
		The corresponding prices are shown in Figure~\ref{fig:stock_price_yf.eps}. 
		Beginning with an initial account value $V(0) = \$1$, we compare our sliding window approach with the classical log-optimal portfolio.

		\begin{figure}[h!]
			\centering
			\includegraphics[width=1\linewidth]{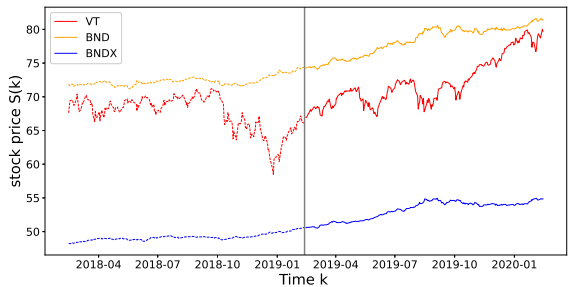}
			\caption{Daily Stock Prices for VT, BND, and BNDX}
			\label{fig:stock_price_yf.eps}
		\end{figure}

		Specifically, to obtain the classical log-optimal portfolio~$K^*$, we solve Problem~(\ref{problem: log-optimal sliding}) for the first year of the data from August~14, 2018 to February~13,~2019.  
		The corresponding optimal $K^*$ is given by 
		$
		K^* = e_3 := [0 \; 0 \;1]^T,
		$
		which suggests that one  should invest all of available funds on BNDX. 
		The out-of-sample performance metrics under classical log-optimal portfolio are summarized in Table~\ref{table:Performance Metrics under Classical Log-Optimal Portfolio}.
		The corresponding account value trajectory is shown in Figure~\ref{fig:value_compare_yf_SLWA.eps} with a bold solid line in black color.
		From the figure, we see that at the terminal stage,~$V(N) \approx 1.0849$ with a maximum percentage drawdown about~$d^*=2.21\%$. 


		\medskip
		\begin{table}[h!] 
			\centering
			\caption{Performance Metrics under Classical Log-Optimal Portfolio}
			\begin{tabular}{ | c | c | } 
				\hline
				\multicolumn{2}{c}{Classical log-optimal portfolio $K^*$}  \\ 
				\hline\hline
				Maximum percentage drawdown $d^*$ & $2.21$\% \\ 
				\hline
				Cumulative rate of return $\frac{V(N) - V_0}{V_0}$ & $8.49$\% \\ 
				\hline
				Realized Log-Growth $\log \frac{V(N)}{V(0)}$ & $8.15$\% \\
				\hline
				volatility (Annualized) $\sigma$ & $2.85$\% \\
				\hline
				Sharpe ratio $\sqrt{N} \cdot SR$ & $2.87$ \\ 
				\hline
				Running Times (secs) & $3.17$  \\
				\hline
			\end{tabular}
			\label{table:Performance Metrics under Classical Log-Optimal Portfolio}
		\end{table}

		\medskip
		\subsection{Log-Optimal Portfolio with Sliding Window Approach}\label{subsection: Kelly optimal portfolio with sliding window approach}
		With various window sizes $M = 5, 10, 30, 60,$ and $100$, we calculate the  log-optimal portfolio weight $K^*(k)$ via the sliding window approach mentioned in Section~\ref{Section: Sliding Window Approach}.
		Having obtained $K^*(k)$, we then implement it on the next trading stage $k+1$; see Figure~\ref{fig:value_compare_yf_SLWA.eps} for an illustration. 
		According to the figure, we find that a relatively small~$M$; e.g., $M=5$ or $M=10$, appears to lead to a higher cumulative return.
		Instead of investing on~BNDX only as suggested by the classical log-optimal portfolio with a constant weight, the sliding window approach yields a time-varying weight; e.g., see Figure~\ref{fig:K_value_compare_yf_SLWA.eps}.
		

		
		\begin{figure}[h!]
			\centering
			\includegraphics[width=1\linewidth]{"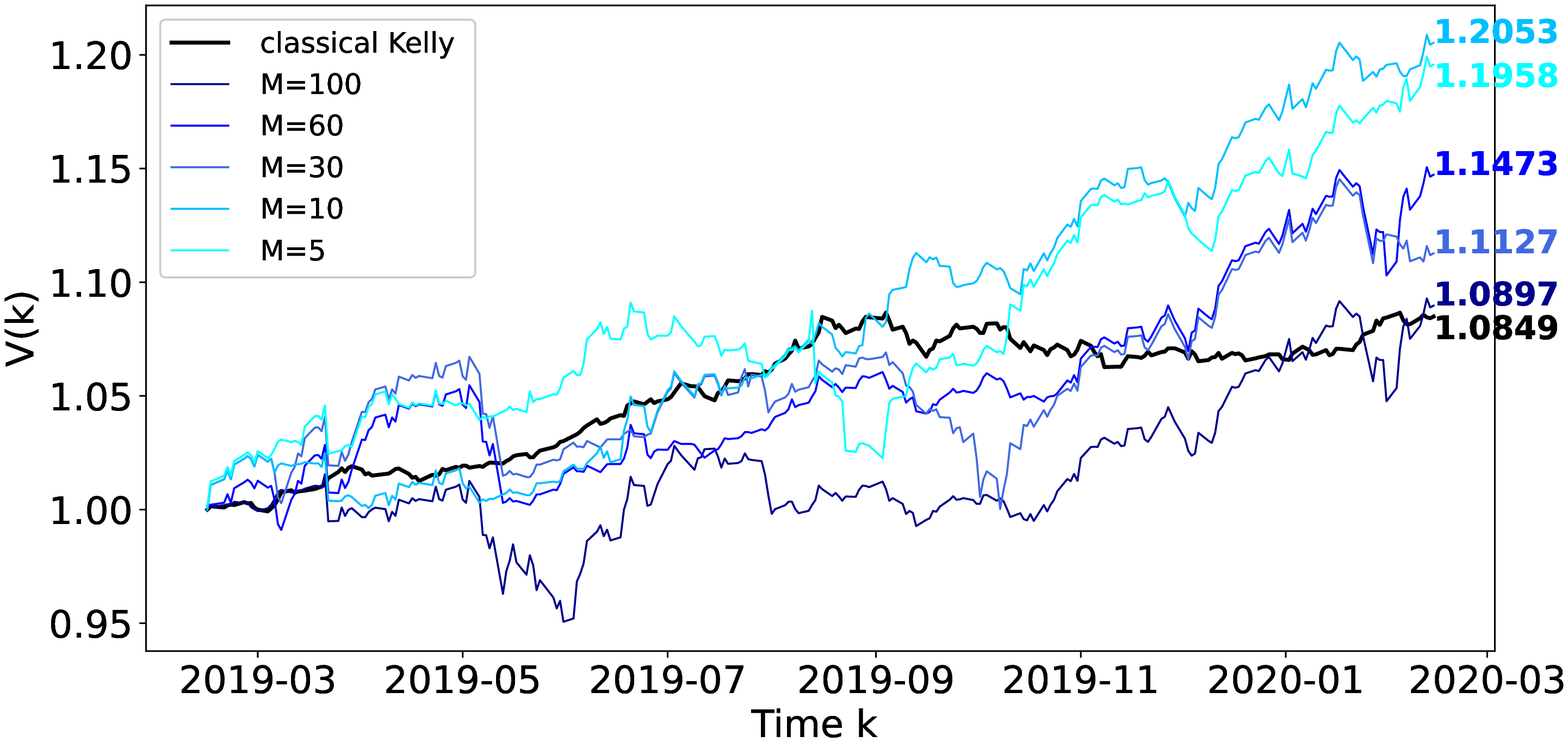"}
			\caption{Classical Log-Optimal Portfolio Versus Sliding Window Approach with  Sizes $M = 5, 10, 30, 60,$ and~$100$}
			\label{fig:value_compare_yf_SLWA.eps}
		\end{figure}

		
		\begin{figure}[h!]
			\centering
			\includegraphics[width=1\linewidth]{"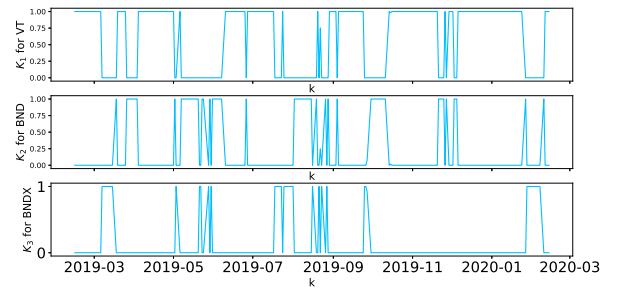"}
			\caption{Time-Varying Portfolio Weights $K(k)$ with size $M = 10$ for the Three-Asset Portfolio}
			\label{fig:K_value_compare_yf_SLWA.eps}
		\end{figure}

		Various performance metrics are summarized in Table~\ref{table:Performance Metrics Under Sliding Window Approach}.
		It is worth noting that one can view the window size $M$ as a new \textit{design} variable in the following sense: One seeks an optimal size $M^*$ that gives the largest return or lowest volatility. In this example, $M=10$ appears to be the best choice in terms of various performance metrics such as Sharpe ratio.

		
		\begin{table*}[h!]
			\centering
			\caption{Performance Metrics Under Sliding Window Approach}
			\begin{tabular}{ | c | c | c | c | c | c | } 
				\hline
				\multicolumn{6}{c}{Log-Optimal Portfolio with Sliding Window Approach}  \\ 
				\hline\hline
				Sliding window sizes $M$  & $100$ & $60$ & $30$ & $10$ & $5$ \\ 
				\hline
				Maximum percentage drawdown $d^*$ & $6.39\%$ & $4.99\%$ & $6.44\%$ & $2.35\%$ & $6.27\%$ \\ 
				\hline
				Cumulative rate of return $\frac{V(N) - V_0}{V_0}$ & $8.80\%$ & $14.53\%$ & $11.29\%$ & $20.55\%$ & $19.40\%$ \\ 
				\hline
				Realized Log-Growth $\log \frac{V(N)}{V(0)}$ & $8.43\%$ & $13.56\%$ & $10.70\%$ & $18.69\%$ & $17.73\%$ \\
				\hline
				Volatility (Annualized) $\sigma$ & $8.36\%$ & $7.70\%$ & $8.30\%$ & $6.30\%$ & $7.54\%$ \\
				\hline
				Sharpe ratio (Annualized) $\sqrt{N}\cdot SR$ & $1.050$ & $1.801$ & $1.331$ & $2.998$ & $2.389$ \\ 
				\hline
				Running Times (secs) & $3.614$ & $6.277$ & $3.325$ & $3.187$ & $2.880$ \\
				\hline
			\end{tabular}
			\label{table:Performance Metrics Under Sliding Window Approach}
		\end{table*}

	\end{example}
	
	\medskip
	\begin{example}[A Hypothetical Case]
		Here we consider the same setting but with the \textit{hypothetical flipped upside-down prices}, which corresponds to a \textit{bear} market situation.\footnote{A bear market refers to as the market in which prices are falling or are expected to fall.}
		Using the sliding window algorithm in Section~\ref{Section: Sliding Window Approach} to test for the flipped upside-down case.
		Figure~\ref{fig:stock_price_yf_up.eps} shows the hypothetical prices of all these three assets (VT, BND, BNDX) from February 14,~2018 to February~14, 2020.
		To obtain a classical log-optimal portfolio $K^*$, we solve the log-optimal portfolio problem for one year ago from February~14,~2018 to February~13, 2019, and apply optimal~$K^*$ into the next year from February 14, 2019 to February 14, 2020. 
		In this case,  corresponding log-optimal portfolio $K^*$ is given by~$
		K^* = e_3 := [0 \; 0 \;1]^T,
		$
		which suggests one  should invest all of available money on asset BNDX. 
		The corresponding account value trajectory is shown in Figure~\ref{fig:value_compare_yf_SLWA_upsidedown.eps} in a bold solid line with black color.
		Other performance metrics are summarized in Table~\ref{table:Performance Metrics of Classical Log-Optimal Portfolio (Hypothetical Case)}.
		
		\begin{table}[h!] 
			\centering
			\caption{Performance Metrics of Classical Log-Optimal Portfolio (Hypothetical Case)}
			\begin{tabular}{ | c | c | } 
				\hline
				\multicolumn{2}{c}{Classical log-optimal portfolio $K^*$}  \\ 
				\hline\hline
				Maximum percentage drawdown $d^*$ & $4.81\%$ \\ 
				\hline
				Cumulative rate of return $\frac{V(N) - V_0}{V_0}$ & $-4.73\%$ \\ 
				\hline
				Realized Log-Growth $\log \frac{V(N)}{V(0)}$ & $-4.84\%$ \\
				\hline
				Volatility (Annualized) $\sigma$ & $1.89\%$ \\
				\hline
				Sharpe ratio $\sqrt{N} \cdot SR$ & $-2.56$ \\ 
				\hline
				Running Times (secs) & $4.36$  \\
				\hline
			\end{tabular}
			\label{table:Performance Metrics of Classical Log-Optimal Portfolio (Hypothetical Case)}
		\end{table}
		
		Similar to Example~\ref{example: three asset portfolio}, we again implement Algorithm~\ref{algorithm: sliding window} with various window sizes $M = 5, 10, 30, 60$, and $100$ days. 
		The corresponding account value trajectories are shown in Figure~\ref{fig:value_compare_yf_SLWA_upsidedown.eps}. 
		Consistent with our theory, the resulting strategy yields a time-varying portfolio weight; see Figure~\ref{fig:K_value_compare_yf_SLWA_upsidedown.eps} for an example of this fact with size $M = 10$.
		
		\begin{figure}[h!]
			\centering
			\includegraphics[width=1\linewidth]{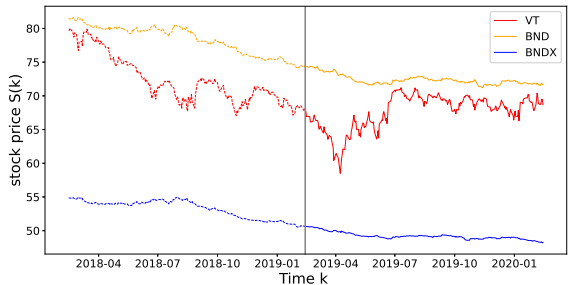}
			\caption{Hypothetical Daily Stock Prices for VT, BND, and BNDX}
			\label{fig:stock_price_yf_up.eps}
		\end{figure}

		
		\begin{figure}[h!]
			\centering
			\includegraphics[width=1\linewidth]{"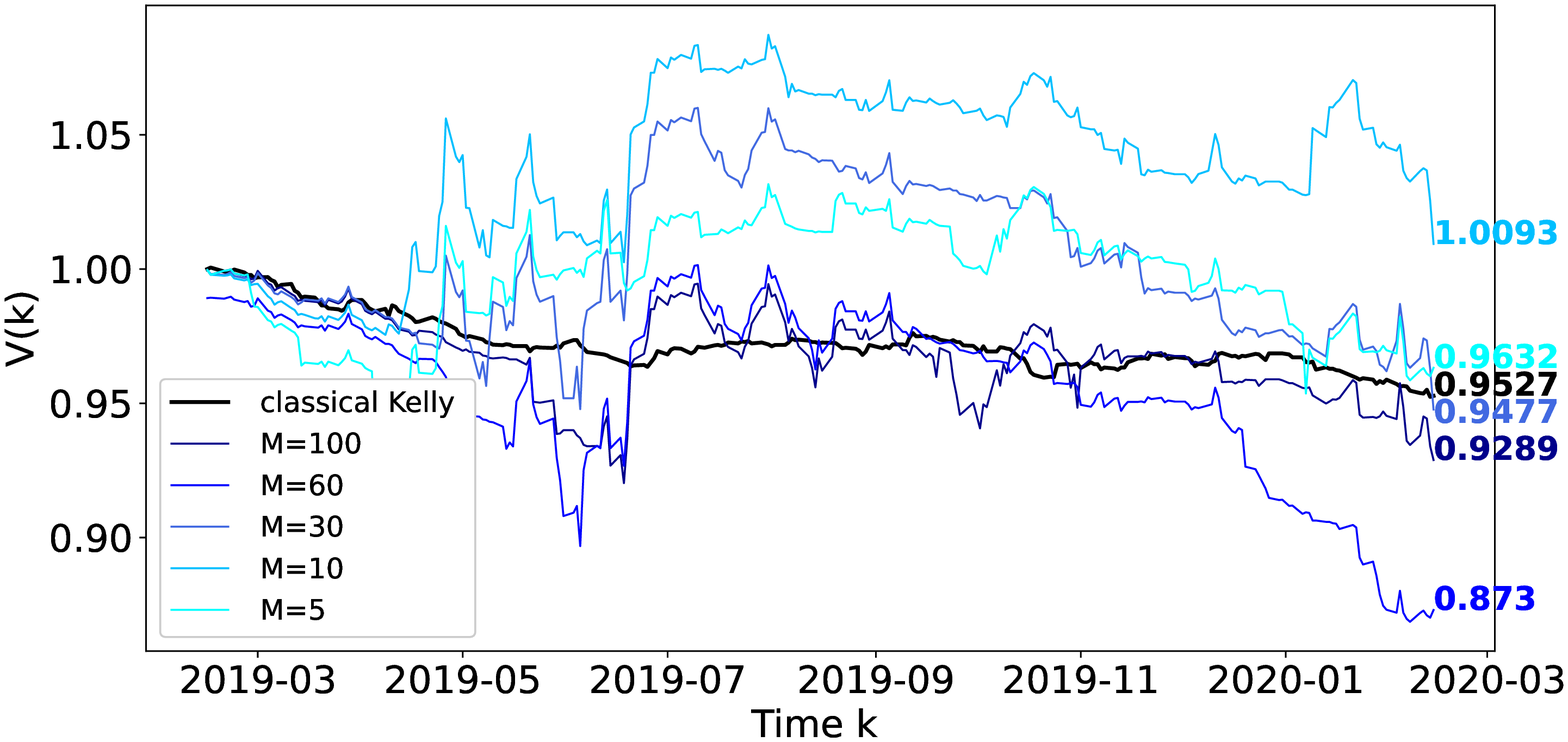"}
			\caption{Classical Log-Optimal Portfolio Versus Sliding Window Approach with  Sizes $M = 5, 10, 30, 60,$ and~$100$ (Hypothetical Case)}
			\label{fig:value_compare_yf_SLWA_upsidedown.eps}
		\end{figure}
		
		
		\begin{figure}[h!]
			\centering
			\includegraphics[width=1\linewidth]{"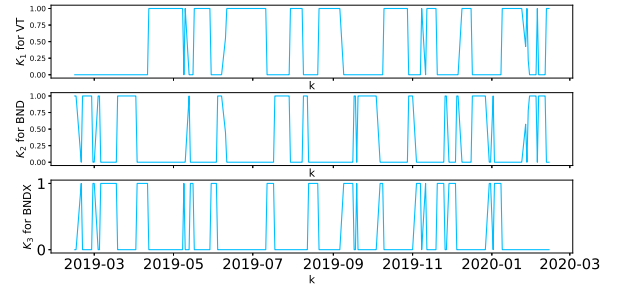"}
			\caption{Time-Varying Portfolio Weights $K(k)$ with size $M = 10$ (Hypothetical Case)}
			\label{fig:K_value_compare_yf_SLWA_upsidedown.eps}
		\end{figure}
		
		Even if in this (hypothetical) {bear} market, the sliding window approach appears to maintain the account value without falling too much. 
		In Table~\ref{table : Performance Metrics Under Sliding Window Approach (Hypothetical Case)}, we see that the annualized rate of return in~$M = 10$ is even surprisingly positive. 


		\begin{table*}[h!]
			\centering
			\caption{Performance Metrics Under Sliding Window Approach (Hypothetical Case)}
			\begin{tabular}{ | c | c | c | c | c | c | } 
				\hline
				\multicolumn{6}{c}{Log-Optimal Portfolio with Sliding Window Approach}  \\ 
				\hline\hline
				Sliding window sizes $M$  & $100$ & $60$ & $30$ & $10$ & $5$ \\ 
				\hline
				Maximum percentage drawdown $d^*$ & $7.96\%$ & $13.27\%$ & $10.60\%$ & $7.16\%$ & $7.56\%$ \\ 
				\hline
				Cumulative rate of return $\frac{V(N) - V_0}{V_0}$ & $-7.08\%$ & $-11.75\%$ & $-5.20\%$ & $0.97\%$ & $-3.64\%$ \\ 
				\hline
				Realized Log-Growth $\log \frac{V(N)}{V(0)}$ & $-7.34\%$ & $-12.49\%$ & $-5.34\%$ & $0.97\%$ & $-3.71\%$ \\
				\hline
				Volatility (Annualized) $\sigma$ & $8.36\%$ & $8.81\%$ & $10.28\%$ & $9.62\%$ & $9.97\%$ \\
				\hline
				Sharpe ratio (Annualized) $\sqrt{N}\cdot SR$ & $-0.858$ & $-1.375$ & $-0.468$ & $0.148$ & $-0.322$ \\ 
				\hline
				Running Times (secs) & $4.826$ & $4.087$ & $3.909$ & $3.580$ & $3.431$ \\
				\hline
			\end{tabular}
			\label{table : Performance Metrics Under Sliding Window Approach (Hypothetical Case)}
		\end{table*}
	\end{example}

	\section{Concluding Remarks} \label{section: Concluding Remarks}
	In this paper, we studied a sliding window approach for solving a data-driven log-optimal portfolio problem.
	In contrast to the classical log-optimal portfolio with a  constant weight, our approach  leads to a time-varying weight, which generalizes the classical case. 
	We show, by example, that our approach may be potentially superior to that with classical log-optimal portfolio in terms of the cumulative account value.
	To close this brief conclusion section, we provide two possible research directions. 
	
	
	
	\textit{Optimal Window Sizes $M$. }
	As mentioned previously in Section~\ref{section: Empirical Sutdies}, the window sizes $M$ can be viewed as a new design variable. 
	Hence, determining an ``optimal" size would be of interest to pursue further.
	It is also worth mentioning that the optimization using a sliding window approach with applying the resulting optimal weights to the next trading day is indeed closely related to the idea of model predictive control; e.g., see~\cite{mayne2000constrained} and \cite{rawlings2017model}. Hence, pursuing this direction might be also fruitful.

	
	\textit{Computational Complexity Issues.}
	While we solve a data-driven log-optimal portfolio problem via a sliding window approach and report  corresponding running times for the algorithm, we did not address the computational complexity of the proposed algorithm.
	Hence, it would be of great interest to analyze the time computational complexities. 
	
	
	

	%
	
	\bibliography{ifacconf}             
	
\end{document}